\documentclass[12pt]{article}
\usepackage{graphicx
,color,xspace,hyperref}
\usepackage{amsmath,amssymb}
\usepackage[numbers,sort&compress,square]{natbib}

\newcommand{\fns}{\footnotesize}

\title{Emergent physics on Mach's  principle and the rotating vacuum}
\author{ 
 G. Jannes$^{1}$ and G.E. Volovik$^{2,3}$
\\
{\fns
$^{1}$ Modelling \& Numerical Simulation Group, Universidad Carlos III de Madrid, }\\
{\fns Avda. de la Universidad 30, 28911 Legan\'{e}s, Spain}
\\
{\fns $^{2}$ Low Temperature Laboratory, Aalto University,  }\\
{\fns P.O. Box 15100, FI-00076 Aalto, Finland}
\\
{\fns $^{3}$ Landau Institute for Theoretical Physics RAS, Kosygina 2,
119334 Moscow, Russia}
}
\begin{document}
\maketitle

\abstract{Mach's principle applied to rotation can be correct if one takes into account the rotation of the quantum vacuum together with the Universe. Whether one can detect the rotation of the vacuum or not depends on its properties.  If the vacuum is fully relativistic at all scales, Mach's principle should work and one cannot distinguish the rotation: in the rotating Universe+vacuum, the co-rotating bucket will have a flat surface (not concave). 
However, if there are ``quantum gravity''  effects which violate Lorentz invariance at high energy, then the rotation will become observable. 
This is demonstrated by analogy in  condensed-matter  systems, which consist of two subsystems: superfluid background (analog of vacuum) and ``relativistic'' excitations (analog of matter). For the low-energy (long-wavelength) observer the rotation of the vacuum is not observable. In the rotating frame, the ``relativistic'' quasiparticles feel the background as a Minkowski vacuum, i.e. they do not feel the rotation.  Mach's idea of the relativity of rotational motion does indeed work for them.  But rotation becomes observable by high-energy observers, who can see the quantum gravity  effects.}


\section{Introduction}
Mach's principle is one of the iconic principles underlying General Relativity (GR). The name ``Mach's principle'' was coined by Einstein for the general inspiration that he found in Mach's works on mechanics~\cite{Mach}, even though the principle itself was never formulated succinctly by Mach himself. In Einstein's words~\cite{Einstein} (see also~\cite{Barbour-book} p.67 and 186): ``Mach's principle: the G[ravitational] field is {\it completely} determined by the masses of the bodies. Since mass and energy are identical ... and formally described by the symmetric energy tensor ($T_{\mu\nu}$), this therefore entails that the G-field be conditioned and determined by the energy tensor.'' In a footnote he adds: ``I have chosen the name 'Mach's principle' because this principle has the significance of a generalization of Mach's requirement that inertia should be derived from an interaction of bodies.'' However, it is unlikely that Mach would have agreed with Einstein's definition~\cite{Barbour-book}. Also, Einstein himself repudiated Mach's principle later in his life after giving up his attempts to formulate GR in a fully Machian sense (see e.g.~\cite{Barbour-book} p.80, 186).

In very general terms, Mach's principle states that the matter distribution of the universe determines its geometry, and that inertial forces are a consequence of this distribution. General relativity seems to conform to this statement, at least in spirit. However, when translating it into concrete requirements, these are often violated by general relativity, or at least some of its solutions. For example, \cite{Bondi:1996md} lists 10 possible formulations of Mach's principle (``Mach1-Mach10'') based on the experimental observation (Mach0) that ``the universe, as represented by the average motion of distant galaxies does not appear to rotate relative to local inertial frames.'' None of Mach1-10 is unambiguously satisfied by GR. As another example, the book~\cite{Barbour-book} contains an index of all 
the possible formulations of Mach's principle grouped into no less than 21 categories, and is largely devoted to discussing their meaning, status and relevance. And the controversy still goes on, see e.g.~\cite{Hartman:2003va,Bhadra:2007pg,Hartman:2008dj} and \cite{Mashhoon:2011qt,Barbour:2011ku,Mashhoon:2011kw}. Other useful references for the historical origin of Mach's principle and its relevance for modern-day physics include \cite{Lichtenegger:2004re} and \cite{Barbour:2010dp}, while a (non-exhaustive) sample of recent articles inspired by Mach's principle includes~\cite{Fearn:2014tja,Lasukov:2014bva,Ghose:2014ela,Telkamp:2014pua,Essen:2013wza,Veto:2013dca,Essen:2012kf}, with several more related to Shape Dynamics (a ``relational'' reformulation of GR loosely inspired by Mach's principle, see e.g.~\cite{Koslowski:2015uda} and references therein). 

Some brief examples might serve to illustrate the delicacy of the question. In Einstein's understanding, an isolated object in an otherwise empty universe should not possess any inertial properties. But this is clearly violated by the Minkowski solution. A cosmological constant would also violate Mach's principle, at least in the strict sense in which the geometry should be determined completely by the mass distribution. Perhaps this can be solved by taking the vacuum energy into account. But the question becomes really delicate when looking at rotating reference frames. These have from the start been at the heart of the debate between absolute and relative motion, as embodied by Newton's bucket argument and Mach's criticism of it, and formed one of the main motivations for Einstein to generalize the special theory of relativity. 
However, GR is probably not fully satisfactory in this respect. Because of the relativity of motion, one could expect a rotation of a local observer w.r.t.\ the rest of the universe to be equivalent to an overall rotation of the universe about the observer. But this idea leads to several difficulties within standard GR (and in fact formed an important source of inspiration for Brans-Dicke theory~\cite{Brans:1961sx}). Also, in the absence of an absolute frame of reference, a global state of motion of the universe---and in particular, a global rotation---should be undetectable (perhaps even meaningless). At first glance, this seems to be violated by the G\"odel solution of GR, which describes a universe with a uniform non-zero rotation in the whole spacetime~\cite{Goedel:1949}. 

All in all, in spite of Barbour's insistence that GR is ``a perfectly Machian theory'', or at least ``as Machian as one could reasonably hope to make any theory'' [\cite{Barbour-book} p.214], it is probably safe to say that GR does not deal with the relativity of rotation in a fully satisfactory way. Perhaps Mach's principle needs to be (partially) given up, or on the contrary GR needs to be partially modified or at least reformulated, or certain restrictions should be imposed on physically acceptable GR solutions, such as in Wheeler's initial-value approach, or Sciama's and Brans-Dicke theory. Another possibility is that Mach's principle will become fully clear only at the level of a quantum theory of gravity~\cite{Barbour-book}.

In this work, we do not pretend to tackle Mach's principle in all its aspects, but restrict ourselves to the question whether it makes sense to speak of an overall rotational motion of the universe, and whether a local observer would be able to detect such a hypothetical overall rotational motion. We will argue that if matter and gravitation as we experience them emerge in the low-energy limit from a quantum vacuum, in a sense intimately related to what occurs in certain condensed-matter systems, then Mach's principle with respect to rotation (in the sense of the undetectability of an overall rotation of the universe) is correct. Two key points that we will stress are, first: that one needs to take the vacuum, and in particular the rotation of the vacuum, into account; and second: that the detectability of rotation depends on the observer, whether he is a low-energy (internal) or a high-energy observer.

The structure of this work is as follows. In Sec.~\ref{S:FluidMetric}~~  the effective metrics emerging in moving superfluids (fluid metric) are discussed, where Mach's principle for rotation is valid for long-wavelength observers (internal observers). For high-energy observers with access to quantum gravity effects, however, the rotation of the vacuum becomes detectable, as discussed in Sec.~\ref{S:caveats}. In Sec.~\ref{S:SuperfluidModel}~ the two-fluid model of the relativistic quantum vacuum is considered, in which the global rotation is not observable. 
We draw some general conclusions in Sec.~\ref{S:conclusions}

\section{Fluid metric and analogue Mach principle}
\label{S:FluidMetric}

We start from the simplest example of the fluid metric:
\begin{equation}
 ds^2 = - dt^2 + \frac{1}{c^2}(d{\bf r} - {\bf v}({\bf r})\, dt)^2 \,.
\label{FluidMetric}
\end{equation}
This metric is valid for long-wavelength excitations with linear spectrum \mbox{$E = cp$} propagating in a background medium (the analog of the quantum vacuum). If the background (vacuum) is moving  with velocity 
${\bf v}(\equiv {\bf v}_{\rm vacuum})$, the spectrum of excitations is Doppler shifted, $E = cp+{\bf p}\cdot {\bf v}$. This
leads to the following contravariant metric:
\begin{equation}
0=\left(E-{\bf p}\cdot {\bf v} \right)^2  -c^2p^2\equiv - g^{\mu\nu}p_\mu p_\nu
\,,
\label{Contravariant}  
\end{equation}
which corresponds to the interval (\ref{FluidMetric}).
 
``Relativistic'' excitations can for example be sound waves (phonons) in a moving fluid or superfluid ($c$ then represents the speed of sound, $c=c_{\rm sound}$,  and the corresponding metric is called the acoustic metric)~\cite{Unruh:1980cg,Barcelo:2005fc,Volovik:2006cz}; or surface waves (ripplons) on a moving liquid or at the interface between two liquids, which in the shallow-depth limit have linear spectrum~\cite{Schutzhold:2002rf,Volovik:2005ga,Rousseaux:2007is,Weinfurtner:2010nu,Jannes:2010sa}.

The  fluid metric for black hole  at the end of the gravitational collapse is the Painleve-Gullstrand metric
\cite{Painleve}, which corresponds to the radial flow field in the form:
   \begin{equation}
 {\bf v}({\bf r})=v(r)\hat{\bf r} ~~,~~v(r)= -c\sqrt{\frac{r_H}{r}}\,,
\label{Schwarzschild}
\end{equation}
where $r_H$ is the radius of lack hole horizon.
The de Sitter spacetime is also characterized by the radial velocity field
\begin{equation}
{\bf v}({\bf r})=v(r)\hat{\bf r} ~~,~~v(r) = c \frac{r}{r_H} \,,
\label{FRWfluid}
\end{equation}
where $r_H$ is the radius of cosmological horizon. We are interested in the fluid metric 
with $\mathbf{v}({\bf r})=\mathbf{\Omega}   \times \mathbf{r}$.

Up to this point, the key motivation to think of superfluids is that, for a meaningful discussion of the observability of a rotating state of the background or vacuum, it is necessary to imagine the construction of measurement devices. It is therefore not sufficient to have sound (or surface/interface) waves: one needs stable fermionic (quasi)particles~\cite{Barcelo:2007iu}. ``Relativistic'' fermionic quasiparticles appear in the topological Weyl superfluids ~\cite{Volovik-book2003},  in graphene \cite{Katsbook}, and in Weyl and Dirac semimetals \cite{Burkov2011}, where in the vicinity of the Weyl or Dirac point the spectrum becomes linear, and these quasiparticles even mimic many properties of the particles of the standard model ~\cite{Volovik-book2003} (see also~\cite{raul-2014}). The quasiparticles in the Weyl superfluid $^3$He-A have an anisotropic linear spectrum, and Eq.(\ref{FluidMetric}) is generalized to
\begin{equation}\label{FluidMetricA-phase}
 ds^2 = - dt^2 +g_{ij}(dx^i- v^i({\bf r})\, dt) (dx^j- v^j({\bf r})\, dt)~,
\end{equation}
where
\begin{equation}
 g_{ij}=\frac{1}{c_\parallel^2} \hat l^i \hat l^j + \frac{1}{c_\perp^2} (\delta^{ij}-\hat l^i \hat l^j)~.
\end{equation}
Here, $c_\parallel$ and $c_\perp$ are the propagation speeds of the massless Weyl quasiparticles (the ``speeds of light'') in the direction parallel or perpendicular to the anisotropy axis $\hat{\bf l}$ (the orientation of the Cooper pairs angular momentum). 

When the background (vacuum) is rotating with a velocity $\mathbf{v} =\mathbf{\Omega}   \times \mathbf{r}$,  the line element in Eq.\eqref{FluidMetric} becomes (in cylindrical coordinates)
\begin{eqnarray}
 ds^2=  -(1 - \frac{r^2\Omega^2}{c^2})dt^2
 \nonumber
 \\
+\frac{1}{c^2}(-2r^2\Omega \,d\phi \,dt + dr^2 + r^2d\phi^2  +dz^2).
\label{RotMetric}
\end{eqnarray}
For $^3$He-A, the anisotropic metric (\ref{FluidMetricA-phase}), with $\mathbf{v} =\mathbf{\Omega}   \times \mathbf{r}$ and $\hat {\bf z}=\hat{\mathbf{\Omega}} =\hat {\bf l}$ gives
\begin{eqnarray}
ds^2=-(1 - \frac{r^2\Omega^2}{c_\perp^2})dt^2
 \nonumber
 \\
 +\frac{1}{c_\perp^2}(-2r^2\Omega \,d\phi \,dt + dr^2 + r^2\,d\phi^2)+\frac{1}{c_\parallel^2}dz^2.
\end{eqnarray}

The speed of light anisotropy can be absorbed by a coordinate rescaling (from the laboratory point of view, $c_\perp \ll c_\parallel$ in $^3$He-A), and in any case we are interested in what happens in the plane of rotation. So we can forget the $z$-coordinate and focus on Eq.~(\ref{RotMetric}), as long as we remember to interpret $c$ correctly. In particular, one should bear in mind that the acoustic waves are associated to excitations of the constituent atoms (and hence of the quasiparticle vacuum), not of the quasiparticles themselves. So from the point of view of the effective gravity of the quasiparticles, the speed of sound is an inaccessible, ultra-high-energy phenomenon. And indeed, in $^3$He-A, the speed of sound $c_{\rm sound}$ is decoupled from the quasiparticle ``speed of light'' $c_\perp$, with  $c_{\rm sound} \gg c_\perp$.

Coming back to the rotating metric Eq.~(\ref{RotMetric}), the coordinate transformation 
\begin{align}
r'&=r\\
z'&=z\\
\phi'&=\phi - \Omega t
\end{align}
reduces this to the standard Minkowski line element (in cylindrical coordinates)
\begin{equation}
 ds^2=-\,dt^2+\frac{1}{c^2}(dr^2+r^2\,d\phi^2+dz^2) \,.
\end{equation}

In other words, the excitations themselves (the sound or surface waves, or the fermionic quasiparticles of the Weyl system), who are coupled to the rotating background, experience a Minkowski spacetime. They do not ``see'' the rotation of the background fluid  but propagate in straight trajectories (in the rotating frame). One could say that the quasiparticles obey an ``analogue Mach principle'', since they are insensitive to the overall rotation of the background, i.e. of the vacuum of their effective universe. 

This simple reasoning also hints that in an emergent scenario, in which matter together with spacetime arise as excitations of an underlying ``ether''-like vacuum, Mach's principle could be realised in a natural way.
There are several caveats to this argument though, which may lead to observability of the vacuum rotation 
by quasiparticles (in other words, by an internal observer, who uses quasiparticles for measurement). 

\section{Caveats}
\label{S:caveats}
{\it Dispersion and dissipation}:
First, the metric description for sound, surface waves and quasiparticle excitations is valid only in the long-wavelength limit. For shorter wavelengths, the spectrum of excitations becomes nonlinear, the emergent Lorentz invariance is violated and the absolute reference frame associated to the vacuum becomes observable. 
E.g., the phononic dispersion relation in a Bose-Einstein condensate can be written in the Bogoliubov form
\begin{eqnarray}\label{dispersion_xi}
\omega^2= c^2k^2 + \frac{1}{4} c^2 \xi^2 k^4~
\end{eqnarray}
in terms of the co-moving frequency and wavenumber $\omega$ and $k$, the (long-wavelength) co-moving speed of sound $c$, and the healing length of the condensate $\xi \equiv \hbar / (mc)$ (with $m$ the mass of the constituent bosons). The equivalent expression in the laboratory frame is obtained through the Doppler shift $\omega({\bf k})=\omega_0(k)+{\bf v}\cdot {\bf k}$. An essentially identical expression is often used as a simple prototype for possible high-frequency Lorentz violations of our universe, see e.g.~\cite{Liberati:2012tb}. 
In an arbitrary, non-concordant reference frame, however, the transformation of the expression \eqref{dispersion_xi} becomes more complicated and the Lorentz-violating coefficients become observer-dependent (see e.g.~\cite{Kostelecky:2000mm}).

Dissipation in the nonlinear regime may also occur, which also introduces an absolute reference frame, leading to dispersion such as
\begin{eqnarray}\label{dissipation}
(\omega-{\bf v}\cdot {\bf k})^2= c^2k^2 -i\Gamma \omega,
\end{eqnarray}
with $\Gamma$ the friction parameter ~\cite{Volovik-book2003}. The $\omega$-dependence of the dissipative term (without any Doppler factor) illustrates the influence of the absolute (laboratory) reference frame.

The complication due to dispersion and dissipation comes as no surprise, since it is likely that our spacetime itself is a long-wavelength concept which breaks down at shorter wavelengths. 
Fortunately, our vacuum is Lorentz invariant with great accuracy: the absence of \v{C}erenkov radiation in the vacuum, as well as the non-detection of any other Lorentz-violating effect in the matter or in the gravity sector so far~\cite{Liberati:2013xla}, show that the lower limit of the energy scale at which the violation of Lorentz symmetry may occur (the ``quantum gravity'' scale) is essentially above the Planck energy scale, $E_{\rm Gr}\gg E_{\rm Planck}$.

 {\it Meniscus}:
As a second caveat, in the above we have assumed that the background itself is not affected by the rotating motion. In reality, there are several different channels in which the vacuum is affected by rotation and which may allow us to detect the global rotation of the vacuum.

In liquids, rotation causes centrifugal effects, which lead to the creation of a meniscus. This lies at the essence of Newton's bucket argument. In liquids, this allows internal observer to detect the rotation of the vacuum. The inhomogeneity of fluid density, which varies with distance from the origin of the rotation, makes the effective metric inhomogeneous.  Indeed, the rotational state of motion will lead, according to Bernoulli's law, to pressure variations $\delta p \sim \rho \,v(r)^2$, and since $c_{\rm sound}^2=\partial p/\partial \rho$, to radius-dependent variations in the density
\begin{equation}\label{rho-variations}
\frac{\delta \rho}{\rho} \sim \frac{v(r)^2}{c_{\rm sound}^2},
\end{equation}
or, in terms of the speed of sound (assuming $c=c(\rho)$ and hence $\delta c=\frac{\partial c}{\partial \rho}\delta \rho$):
\begin{equation}\label{c-variations}
\frac{\delta c_{\rm sound}}{c_{\rm sound}} = \alpha \frac{\delta \rho}{\rho} \sim \frac{v(r)^2}{c_{\rm sound}^2},
\end{equation}
since $\alpha=\frac{\partial \ln c}{\partial \ln\rho}$ is a material parameter typically of order unity. 

This will be detected by the internal observer as an inhomogeneity in the gravitational field, although it becomes undetectable in the limit \mbox{$c_{\rm sound}\rightarrow \infty$} (i.e., for incompressible fluids). In other words, the centrifugal effect is important for the acoustic metric, but not for the quasiparticle metric, since  the ``speed of light'' $c_\perp$ is much smaller than $c_{\rm sound}$, as remarked above in Sec.~\ref{S:FluidMetric}. However, one should again bear in mind that $c_{\rm sound}$ is associated to excitations of the constituent atoms (and hence of the quasiparticle vacuum). So, applying the previous observations to the quantum vacuum, we may conclude that any centrifugal effect of vacuum rotation would be associated to a finite compressibility of the quantum vacuum, which is again a ``quantum gravity'' effect, non-accessible to low-energy observers.

The condition $c\ll c_{\rm sound}$ allows us to reach velocities of the background which exceed the effective speed of light. With respect to rotation this means the metric \eqref{RotMetric}  is physical even beyond the ergosurface at $r_\Omega=c/\Omega$, where the linear velocity $v_\phi(r)$ exceeds the speed of light. This does not violate Lorentz invariance, which requires that the velocity of matter (quasiparticles) {\it with respect to the vacuum} cannot exceed the speed of light, 
$|{\bf v}_{\rm matter}-{\bf v}_{\rm vacuum}| \leq c$, while there is no constraint on the velocity of the vacuum itself.   In a similar way the radial coordinate velocity in an expanding Universe exceeds the speed of light beyond the cosmological horizon, see Eq.(\ref{FRWfluid}) for  Sitter Universe.

{\it Vortices}:
In superfluids, another effect of rotation on the structure of the vacuum, which leads to a third possible caveat, becomes manifest. The superfluid may experience the solid-body rotation only on average, because of the quantization of vorticity of the superfluid velocity, $\oint_Cd{\bf r}\cdot {\bf v}= N\kappa$, where  $\kappa$ is the circulation quantum, and $N$ is the number of quanta (number of quantized vortices) within the contour $C$. The equilibrium rotational state of the superfluid with $\left<\mathbf{v} \right>=\mathbf{\Omega}   \times \mathbf{r}$ contains quantized vortex lines with areal density \mbox{$n=N/A=2\Omega/\kappa$}.
If in the quantum vacuum the parameter $\kappa\neq 0$, the rotating state of such vacuum represents a vortex crystal with intervortex distance $l\sim n^{-1/2}$.  This is different from the superfluid model discussed in Refs. \cite{Chapline:2004mu,Chapline:2005hm,Chapline:2009cp}, where the system of quantized vortices produces the G\"odel  metric instead of rigidly rotating vacuum.  

The distance $l$ between the vortices provides another length scale of ``quantum gravity'', which depends on the rotation velocity of the vacuum, and could be detected as an inhomogeneity of the vacuum by observers with access to such quantum gravity scales. For example,  $\kappa$ can be on the order of  $\kappa \sim 
\hbar c^2/E_ {\rm Gr}$, at which Lorentz invariance is violated in the static vacuum (for our present almost Lorentz invariant vacuum $l_{\rm Gr}\ll l_{\rm Planck}$). Then the second scale is
$l=(l_{\rm Gr}r_\Omega)^{1/2}$. 

{\it Discussion}:
These caveats impose some constraints on the realization of Mach's principle in emergent physics, in which matter together with spacetime arise as excitations of an underlying ``ether''-like vacuum.  
The observability of the global rotation depends on quantum gravity effects, i.e. on the microscopic structure of the quantum vacuum. That is why it is possible that for those observers who use long-wavelength quasiparticles for their measurements, the global rotation is not detectable, while it becomes detectable
at higher energies. If, however, the considered ``quantum gravity'' effects are absent, then the Mach principle for rotation becomes valid. For that, the quantum vacuum must be Lorentz invariant for all scales without any restriction on circulation. 

Even if the global rotational state becomes undetectable in the absence of quantum gravity effects, this does not mean that the question of the overall motion of the system is ill-defined: the vacuum has a definite state of motion, but the quasiparticles are unable to detect it. In fact, in a laboratory superfluid in the zero-temperature limit, the vacuum finds its own reference frame, decoupled from the laboratory frame~\cite{Hosio-PRL2011}. From the laboratory point of view, we can determine this reference frame because we have a privileged ``external'' view of the system, whereas the quasiparticles are limited to an internal view, and hence are unable to discern the reference frame of the vacuum.

Next we consider the superfluid model, which incorporates the constraints needed for the Mach principle
to be valid. 

\section{Superfluid model of rotating Universe}
\label{S:SuperfluidModel}

 We assume that the vacuum obeys Galilean physics, while the excitations over the vacuum have a linear spectrum  $E_0(p)=cp$ in the frame comoving with the vacuum. For massless elementary particles, $c$ is the speed of light with respect to the vacuum, for gapless quasiparticles -- excitations of the liquid -- the speed  $c$ is the maximum attainable velocity with respect to the superfluid component.  

In superfluids, the combined dynamics of "vacuum" and "matter" is incorporated  in the Landau-Khalatnikov two-fluid hydrodynamics \cite{Khalatnikov}. The energy density of the liquid in an arbitrary frame is
\begin{equation}
\epsilon= \Lambda+  \frac{1}{2}\rho {\bf v}_s^2 +{\bf P}\cdot{\bf v}_s +\epsilon_{\rm m}~~,~~
\epsilon_{\rm m}=T^{00}=\sum_{\bf p}f({\bf p})E_0(p)\,.
\label{SuperfluidEnergy}
\end{equation}
Here ${\bf v}_s$ is the superfluid velocity, which is the analogue of the velocity of vacuum;  $\epsilon_{\rm m}$ is the energy density of matter in the vacuum comoving frame, i.e. in the frame moving with ${\bf v}_s$; ${\bf P}$ is the momentum density of matter in that frame;  $f({\bf p})$ is the distribution function;  and $\Lambda$ is the energy density of the superfluid vacuum, which is (Galilean) frame-independent. It is the analog of the cosmological constant: the vacuum has an equation of state \mbox{$P_{\rm vac}=-\epsilon_{\rm vac}=-\Lambda$}, while for the massless relativistic matter one has
\mbox{$P_{\rm m}=\frac{1}{3}\epsilon_{\rm m}$}.

Galilean physics is consistent with the observed Lorentz invariance of our vacuum only if the total density of the superfluid is $\rho=0$. This is because $\rho$ is the time component of the 4-vector, and so any Lorentz transformation with a non-zero $\rho$ would break the equivalence of the reference frames. In the two-fluid model of superfluid liquid this means that the densities of the normal and superfluid components of the liquid must cancel each other:
\begin{equation}
 \rho=\rho_s  + \rho_n=0 \,.
\label{TotalDensity}
\end{equation}
In general, the momentum density of the liquid obeys ${\bf j}= \rho {\bf v}_s +{\bf P}=\rho {\bf v}_s + \sum_{\bf p}  {\bf p}f({\bf p})$.\cite{Khalatnikov} 
Since $\rho=0$, the vacuum has zero momentum, and thus is indeed invariant under transformation to a moving frame (i.e. it is invariant under Galilean or Lorentz transformations). 
The momentum is then only carried by matter (``relativistic'' quasiparticles), and can be written 
\begin{equation}
{\bf P}=\sum_{\bf p}  {\bf p}f({\bf p}) =\rho_n ({\bf v}_n-{\bf v}_s)~.
\label{Momentum}
\end{equation}
The last equation is for thermal equilibrium, where for fermions and bosons one has \mbox{$f({\bf p})=[\exp((E({\bf p})-{\bf p}\cdot{\bf v}_n)/T)\pm 1]^{-1}$},  \mbox{$E({\bf p})=cp + {\bf p}\cdot{\bf v}_s$}.
Here the velocity ${\bf v}_n$ of the heat bath of matter is introduced, which in two-fluid hydrodynamics is called the velocity of the normal component of the liquid. 

Combining these observations with the conclusions of the previous section, the Lorentz invariant quantum vacuum can be simulated by a superfluid under the following conditions.
(i) The density of this superfluid must be $\rho=0$. As just discussed, this is required to guarantee the Lorentz invariance of the vacuum. Also the centrifugal force is absent for $\rho=0$. (ii)  Vanishing circulation quantum $\kappa=0$. In this case no periodic vortex states appear under rotation.  (iii) Infinite speed of sound, $c_{\rm sound}=\infty$ (as is the case in $^3$He-A: $c_{\rm sound} \gg c_\perp$). Then there is no observable compressibility of the vacuum.

In this superfluid model in which $\rho_s=-\rho_n < 0$, the superfluid vacuum is unstable towards the growth  of $|{\bf v}_s-{\bf v}_n|$.\footnote{There are other arguments why a superfluid model cannot fully reproduce our macroscopic physics, e.g. related to the hierarchy of scales that would be required to obtain Einstein dynamics, see~\cite{Volovik-book2003,Barcelo:2010vc}. But, as often, studying how far the model can be taken is far more instructive than dismissing it altogether.} Eventually, at $|{\bf v}_s-{\bf v}_n|=c$, a limiting state would be reached, corresponding to the Landau critical velocity.

In spite of the instability, this superfluid model can be used for the discussion of the rotating vacuum, in which both vacuum and matter experience the solid-body rotation, $\mathbf{v}_s\equiv \mathbf{v}_{\rm vacuum} =\mathbf{\Omega}   \times \mathbf{r}$ and $\mathbf{v}_n  \equiv \mathbf{v}_{\rm matter}=\mathbf{\Omega}   \times \mathbf{r}$. In this model all the rotating states 
have the same thermodynamic potential, if the matter is co-rotating with the vacuum. 
The degenerate states corresponding to different values of $\mathbf{\Omega}$ are not distinguishable by the internal observers, since all of them observe the Minkowski metric. So the Mach principle does indeed work in this model of the vacuum.

In fact, for a sufficiently large system (behind the cosmological horizon), the matter component must necessarily co-rotate with the vacuum. Otherwise, for large enough $r> c/\Omega$, one would inevitably exceed the Landau critical velocity $|{\bf v}_s-{\bf v}_n|=c$, at which the vacuum becomes unstable due to the spontaneous production of quasiparticles and vortices. The development of the instability finally results in the states  with $\mathbf{\Omega}_{\rm vacuum}=\mathbf{\Omega}_{\rm matter}$.  Note that in the considered model all such states are still unstable towards a linear increase of 
$|{\bf v}_s-{\bf v}_n|$ due to the negative $\rho_s<0$.

\section{Conclusions}\label{S:conclusions}

From an emergent-physics point of view, Mach's principle with respect to rotating reference frames ultimately lies beyond GR. The essential question is in which reference frame the quantum vacuum is at rest, and whether this can be observed. This is a question for GR plus the quantum vacuum, not for GR only. 

When taking the quantum vacuum into account, Mach's principle works fine for low-energy observers: these are unable to discern between different rotating states of the vacuum. We have illustrated this by focusing on the fluid metric in condensed-matter models, with particular attention to the acoustic and quasiparticle metrics in superfluids. High-energy observers with access to ``quantum gravity'' effects, however, would still be able to detect the overall rotating state of the system. Next, we have examined a superfluid toy-model for the universe where these quantum gravity effects are absent. Although the model itself is unstable and thus not fully realistic, it is useful for the question of rotating reference frames. In particular, it clearly illustrates that, in a sufficiently large universe, the matter component must necessarily co-rotate with the vacuum, and that Mach's principle with respect to rotation indeed is again valid.

\section*{\small Acknowledgements}
{\small
G.J. acknowledges financial support from the Spanish MICINN through project FIS2011-30145-C03-01, and a travel grant from the Magnus Ehrnrooth Foundation.
The work by G.V. has been supported in part by the Academy of Finland
(Centers of Excellence Programme project no.
250280). We thank A. Starobinsky for fruitful discussions.
}


\begin{thebibliography}{99}
\bibitem{Mach} E. Mach, {\it Die Mechanik in ihrer Entwickelung:  historisch-kritisch dargestellt}, Brockhaus, Leipzig (1883).

\bibitem{Einstein}
A. Einstein, 
``Prinzipielles zur allgemeinen Relativit\"atstheorie,''
Ann. Phys. {\bf 360}, 241 (1918)

\bibitem{Barbour-book}
J. Barbour and H. Pfister (eds.), {\it Mach's Principle: From Newton's Bucket to
Quantum Gravity}, Birkhauser, Boston, MA (1995)

\bibitem{Bondi:1996md} 
  H.~Bondi and J.~Samuel,
  ``The Lense-Thirring effect and Mach's principle,''
    Phys.\ Lett.\ A {\bf 228}, 121 (1997)



\bibitem{Hartman:2003va} 
  H.~I.~Hartman and C.~Nissim-Sabat,
  ``On Mach's critique of Newton and Copernicus,''
  Am.\ J.\ Phys.\  {\bf 71}, 1163 (2003)

\bibitem{Bhadra:2007pg} 
  A.~Bhadra and S.~C.~Das,
  ``Comment on 'On Mach's critique of Newton and Copernicus',''
  Am.\ J.\ Phys.\  {\bf 75}, 850 (2007)

\bibitem{Hartman:2008dj} 
  H.~I.~Hartman and C.~Nissim-Sabat,
``Reply in Light of Contemporary Physics to ``Comment on 'On Mach’s critique of Newton and \mbox{Copernicus' '',''}
  Am.\ J.\ Phys.\  {\bf 75}, 854 (2007)

\bibitem{Mashhoon:2011qt} 
  B.~Mashhoon and P.~S.~Wesson,
  ``Mach's Principle and Higher-Dimensional Dynamics,''
  Annalen Phys.\  {\bf 524}, 63 (2012)

\bibitem{Barbour:2011ku} 
  J.~Barbour,
  ``Mach's Principle: A Response to Mashhoon and Wesson's Paper arXiv: 1106.6036,''
  Annalen Phys.\  {\bf 524}, A39 (2012)

\bibitem{Mashhoon:2011kw} 
  B.~Mashhoon and P.~S.~Wesson,
  Annalen Phys.\  {\bf 524}, A44 (2012)

\bibitem{Lichtenegger:2004re} 
  H.~Lichtenegger and B.~Mashhoon,
  ``Mach's principle,'' in: L. Iorio (ed.), {\it The Measurement of Gravitomagnetism: A Challenging Enterprise}, Nova Science, New York (2007), pp. 13-25

\bibitem{Barbour:2010dp} 
  J.~Barbour,
  ``The Definition of Mach's Principle,''
  Found.\ Phys.\  {\bf 40}, 1263 (2010)

\bibitem{Fearn:2014tja} 
  H.~Fearn,
  ``Mach's principle, Action at a Distance and Cosmology,''
  J.\ Mod.\ Phys.\  {\bf 6}, 260 (2015)  

\bibitem{Lasukov:2014bva} 
  V.~V.~Lasukov, E.~Y.~Danilyuk and E.~E.~Ilkin,
  ``Mach`s Principle in the Atomic Formulation and the Classical Potential Vacuum,''
  Russ.\ Phys.\ J.\  {\bf 57}, no. 6, 783 (2014)

\bibitem{Ghose:2014ela} 
  P.~Ghose,
  ``Unification of Gravity and Electromagnetism I: Mach's Principle and Cosmology,''
  arXiv:1408.2403 [gr-qc]
  
\bibitem{Telkamp:2014pua} 
  H.~Telkamp,
  ``A relational approach to the Mach-Einstein question,''
  arXiv:1404.4046 [gr-qc]

\bibitem{Essen:2013wza} 
  H.~Ess\'en,
Journal of Gravity 2014, 415649 (2014)
 
\bibitem{Veto:2013dca} 
  B.~Vet\H{o},
  ``Retarded cosmological gravity and Mach's principle in flat FRW universes,''
  arXiv:1302.4529 [gr-qc]

\bibitem{Essen:2012kf} 
  H.~Ess\'en,
  ``Mechanics, cosmology and Mach's principle,''
  Eur.\ J.\ Phys.\  {\bf 34}, 139 (2013)

\bibitem{Koslowski:2015uda} 
  T.~Koslowski,
  ``The shape dynamics description of gravity,''
  arXiv:1501.03007 [gr-qc]

  
  
  
\bibitem{Brans:1961sx} 
  C.~Brans and R.~H.~Dicke,
  ``Mach's principle and a relativistic theory of gravitation,''
  Phys.\ Rev.\  {\bf 124}, 925 (1961)

\bibitem{Goedel:1949}
  K.~G\"odel, 
  ``An Example of a new type of cosmological solutions of Einstein's field equations of graviation,'' 
  Rev.\ Mod.\ Phys.\  {\bf 21}, 447 (1949)


\bibitem{Unruh:1980cg} 
  W.~G.~Unruh,
  ``Experimental black hole evaporation,''
  Phys.\ Rev.\ Lett.\  {\bf 46}, 1351 (1981)

\bibitem{Barcelo:2005fc} 
  C.~Barcel\'o, S.~Liberati and M.~Visser,
  ``Analogue gravity,''
  Living Rev.\ Rel.\  {\bf 14}, 3 (2011)

\bibitem{Volovik:2006cz} 
  G.~E.~Volovik,
  ``Black-hole and white-hole horizons in superfluids,''
  J.\ Low.\ Temp.\ Phys.\  {\bf 145}, 337 (2006)

\bibitem{Schutzhold:2002rf} 
  R.~Schutzhold and W.~G.~Unruh,
  ``Gravity wave analogs of black holes,''
  Phys.\ Rev.\ D {\bf 66}, 044019 (2002)

\bibitem{Volovik:2005ga} 
  G.~E.~Volovik,
  ``The Hydraulic jump as a white hole,''
  JETP Lett.\  {\bf 82}, 624 (2005)
  [Pisma Zh.\ Eksp.\ Teor.\ Fiz.\  {\bf 82}, 706 (2005)]

\bibitem{Rousseaux:2007is} 
  G.~Rousseaux, C.~Mathis, P.~Maissa, T.~G.~Philbin and U.~Leonhardt,
  ``Observation of negative phase velocity waves in a water tank: A classical analogue to the Hawking effect?,''
  New J.\ Phys.\  {\bf 10}, 053015 (2008)

\bibitem{Weinfurtner:2010nu} 
  S.~Weinfurtner, E.~W.~Tedford, M.~C.~J.~Penrice, W.~G.~Unruh and G.~A.~Lawrence,
  ``Measurement of stimulated Hawking emission in an analogue system,''
  Phys.\ Rev.\ Lett.\  {\bf 106}, 021302 (2011)

\bibitem{Jannes:2010sa} 
  G.~Jannes, R.~Piquet, P.~Maissa, C.~Mathis and G.~Rousseaux,
  ``Experimental demonstration of the supersonic-subsonic bifurcation in the circular jump: A hydrodynamic white hole,''
  Phys.\ Rev.\ E {\bf 83}, 056312 (2011)

\bibitem{Painleve} 
P. Painlev\'e, 
 ``La m\'ecanique classique et la th\'eorie de la relativit\'e, '' 
 C. R. Acad. Sci. (Paris) {\bf 173}, 677 (1921); 
A. Gullstrand,
  ``Allgemeine L\"osung des statischen Eink\"orper-problems in der Einsteinschen Gravitations-theorie,'' 
 Arkiv. Mat. Astron. Fys.  {\bf 16}(8), 1 (1922)

\bibitem{Barcelo:2007iu} 
  C.~Barcel\'o and G.~Jannes,  
  ``A Real Lorentz-FitzGerald contraction,''
  Found.\ Phys.\  {\bf 38}, 191 (2008)


\bibitem{Volovik-book2003} 
G.E. Volovik, 
{\it The Universe in a Helium Droplet}, 
Clarendon Press,  Oxford (2003)

\bibitem{Katsbook}
M.I. Katsnelson, 
{\it Graphene: Carbon in Two Dimensions}, 
Cambridge University Press, Cambridge (2012).

\bibitem{Burkov2011}
A.A. Burkov and L. Balents, 
``Weyl semimetal in a topological insulator multilayer,'' 
Phys. Rev. Lett. {\bf 107}, 127205 (2011);
A.A. Burkov, M.D. Hook, L. Balents,
``Topological nodal semimetals,'' 
Phys. Rev. B {\bf 84}, 235126 (2011).


\bibitem{raul-2014}
C.~Barcel\'o, R.~Carballo, L.~J.~Garay and G.~Jannes,
``Electromagnetism as an emergent phenomenon: a step-by-step guide,'' 
New J.\ Phys.\  {\bf 16}, 123028 (2014)

\bibitem{Liberati:2012tb} 
  S.~Liberati,
  ``Lorentz breaking Effective Field Theory and observational tests,''
  Lect.\ Notes Phys.\  {\bf 870}, 297 (2013)

\bibitem{Kostelecky:2000mm} 
  V.~A.~Kostelecky and R.~Lehnert,
  ``Stability, causality, and Lorentz and CPT violation,''
  Phys.\ Rev.\ D {\bf 63}, 065008 (2001)

\bibitem{Liberati:2013xla} 
  S.~Liberati,
 ``Tests of Lorentz invariance: a 2013 update,''
  Class.\ Quant.\ Grav.\  {\bf 30}, 133001 (2013)

\bibitem{Chapline:2004mu} 
  G.~Chapline and P.~O.~Mazur,
  ``Superfluid picture for rotating space-times,''
  Acta Phys.\ Polon.\ B {\bf 45}, no. 4, 905 (2014)

\bibitem{Chapline:2005hm} 
  G.~Chapline and P.~O.~Mazur,
  ``Tommy gold revisited: why does not the universe rotate?''
  AIP Conf.\ Proc.\  {\bf 822}, 160 (2006)

\bibitem{Chapline:2009cp} 
  G.~Chapline and P.~O.~Mazur,
  ``Superfluidity and Stationary Space-Times,''
  arXiv:0911.2326 [hep-th]

\bibitem{Hosio-PRL2011}
J.J. Hosio, V.B. Eltsov, R. de Graaf, P.J. Heikkinen, R. Hanninen, M. Krusius, V.S. L'vov, G.E. Volovik,
``Superfluid vortex front at T $\rightarrow$ 0: Decoupling from the reference frame,''
 Phys.\ Rev.\ Lett.\ {\bf 107}, 135302 (2011)

\bibitem{Khalatnikov} 
I.~M. Khalatnikov,   
{\it An Introduction to the Theory of Superfluidity},
Benjamin, New York  (1965)

\bibitem{Barcelo:2010vc} 
  C.~Barcel\'o, L.~J.~Garay and G.~Jannes,
  ``Quantum Non-Gravity and Stellar Collapse,''
  Found.\ Phys.\  {\bf 41}, 1532 (2011)


  
 
 \end{thebibliography}
\end{document}